\documentclass[a4paper]{jpconf}
\usepackage{graphicx}
\usepackage{paralist}
\usepackage{xspace}
\usepackage{xcolor}
\usepackage{listings}

\def\CO2 {CO$_2$\ }

\makeatletter
\newlength\cxx@c@height
\newlength\cxx@plus@height
\newcommand\cxx@plus@fontsize{\fontsize{1.6ex}{1.92ex}\selectfont}
\DeclareRobustCommand{\CC}{%
  \settoheight\cxx@c@height{C}%
  \settoheight\cxx@plus@height{\cxx@plus@fontsize+}%
  \advance\cxx@c@height by -\cxx@plus@height%
  \multiply\cxx@c@height by 10% div by 2.3
  \divide\cxx@c@height by 23%
  C\kern-.08em\raise\cxx@c@height\hbox{\cxx@plus@fontsize+\kern-.08em+}%
\xspace}
\makeatother

\newcommand{\clqcd}{CL\kern-.25em\textsuperscript{2}QCD}

\def\splitfirstchar#1#2\sentinel{\textbf{#1}#2}
\newcommand{\fBold}[1]{\splitfirstchar#1\sentinel}
\newcommand{\fig}[1]{Fig.~\ref{fig:#1}}

\newcommand*{\eg}{e.\,g.\@\xspace}
\newcommand*{\ie}{i.\,e.\@\xspace}

\definecolor{MyGreen}{rgb}{0,0.5,0}
\begin{document}

\title{Fast TPC Online Tracking on GPUs and Asynchronous Data Processing in the ALICE HLT to facilitate Online Calibration}

\author{David Rohr, Sergey Gorbunov, Mikolaj Krzewicki, Timo Breitner, Matthias Kretz, Volker Lindenstruth for the ALICE Collaboration}

\address{Frankfurt Institute for Advanced Studies, Ruth-Moufang-Str. 1, 60438 Frankfurt, Germany}

\ead{drohr@cern.ch}

\begin{abstract}
\looseness=-1
ALICE (A Large Heavy Ion Experiment) is one of the four major experiments at the Large Hadron Collider (LHC) at CERN, which is today the most powerful particle accelerator worldwide.
The High Level Trigger (HLT) is an online compute farm of about 200 nodes, which reconstructs events measured by the ALICE detector in real-time.
The HLT uses a custom online data-transport framework to distribute data and workload among the compute nodes.

\looseness=-1
ALICE employs several calibration-sensitive subdetectors, e.g. the TPC (Time Projection Chamber).
For a precise reconstruction, the HLT has to perform the calibration online.
Online-calibration can make certain Offline calibration steps obsolete and can thus speed up Offline analysis.
Looking forward to ALICE Run III starting in 2020, online calibration becomes a necessity.

\looseness=-1
The main detector used for track reconstruction is the TPC.
Reconstructing the trajectories in the TPC is the most compute-intense step during event reconstruction.
Therefore, a fast tracking implementation is of great importance.
Reconstructed TPC tracks build the basis for the calibration making a fast online-tracking mandatory.

We present several components developed for the ALICE High Level Trigger to perform fast event reconstruction and to provide features required for online calibration.

\looseness=-1
As first topic, we present our TPC tracker, which employs GPUs to speed up the processing, and which bases on a Cellular Automaton and on the Kalman filter.
Our TPC tracking algorithm has been successfully used in 2011 and 2012 in the lead-lead and the proton-lead runs.
We have improved it to leverage features of newer GPUs and we have ported it to support OpenCL, CUDA, and CPUs with a single common source code.
This makes us vendor independent.

\looseness=-1
As second topic, we present framework extensions required for online calibration.
The extensions, however, are generic and can be used for other purposes as well.
We have extended the framework to support asynchronous compute chains, which are required for long-running tasks required e.g. for online calibration.
And we describe our method to feed in custom data sources in the data flow.
These can be external parameters like environmental temperature required for calibration and these can also be used to feed back calibration results into the processing chain.

Overall, the work presented in this contribution makes the ALICE HLT ready for online reconstruction and calibration for the LHC Run II starting in 2015.
\end{abstract}

\section{Introduction}

\begin{figure}[h]
\begin{minipage}{15.5pc}
\includegraphics[width=15.5pc]{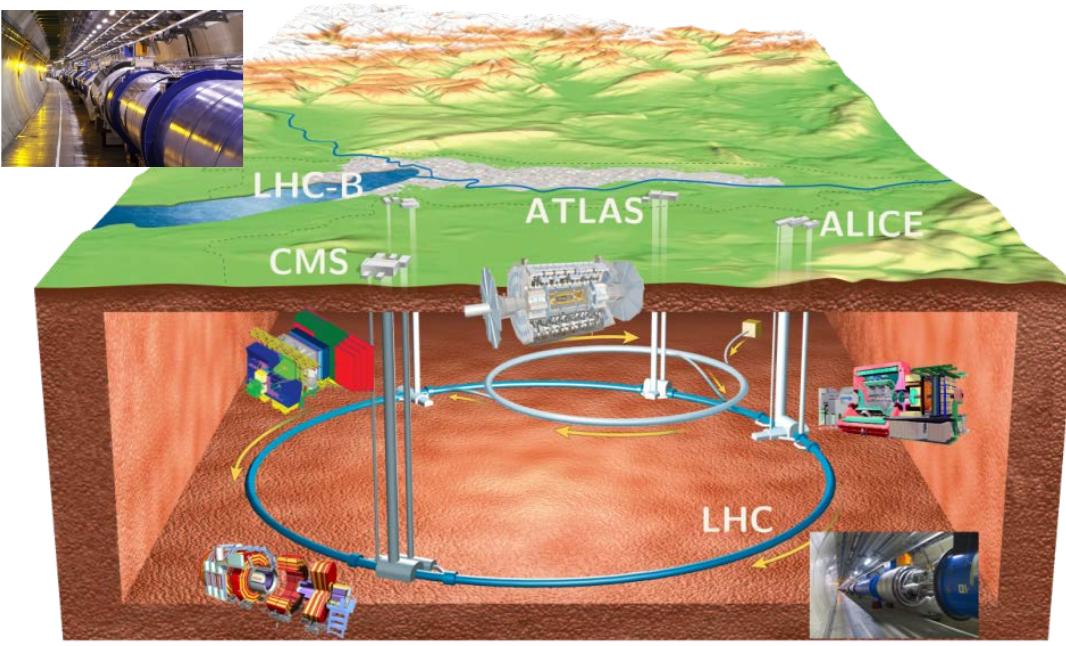}
\caption{The Large Hadron Collider beneath Geneva and its four major experiments.}
\label{fig:lhc}
\end{minipage}\hspace{2pc}%
\begin{minipage}{14.5pc}
\includegraphics[width=14.5pc]{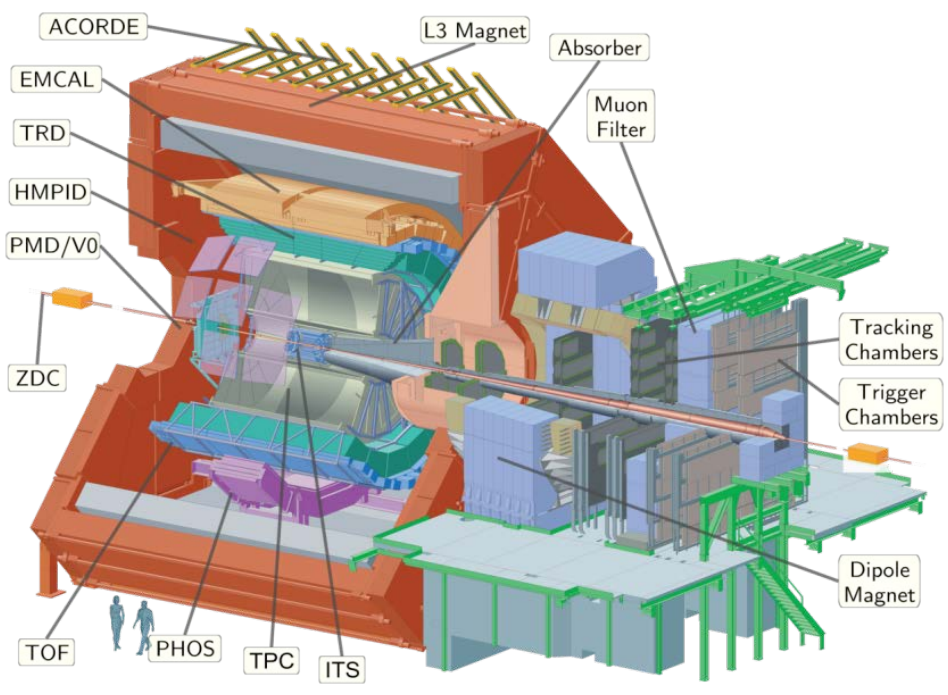}
\caption{The ALICE detector and subdetectors.}
\label{fig:alice}
\end{minipage}
\end{figure}

\looseness=-1
The \fBold{Large} \fBold{Hadron} \fBold{Collider} (\textbf{LHC}, \fig{lhc}) at CERN is today's most powerful particle accelerator.
\fBold{A} \fBold{Large} \fBold{Ion} \fBold{Collider} \fBold{Experiment} (\textbf{ALICE}~\cite{alice_technical-proposal}, \fig{alice}) is a dedicated experiment designed to study heavy ion physics by exploiting the physics potential of lead nucleus collisions at LHC.
The baseline detector design (from inner to outer region) starts with the silicon detector \textbf{ITS} (\fBold{Inner} \fBold{Tracking} \fBold{System}), followed by the cylindrical gas detector \textbf{TPC} (\fBold{Time} \fBold{Projection} \fBold{Chamber}), the \textbf{TRD} (\fBold{Transition} \fBold{Radiation} \fBold{Detector}), particle identification detectors, and calorimeters.

The reconstruction of the recorded events happens at two places.
First, the \fBold{High} \fBold{Level} \fBold{Trigger} (\textbf{HLT}), which is an online compute farm of roughly~200 compute nodes, performs a fast online-event reconstruction in realtime.
The HLT can run trigger-algorithms on the reconstructed event data which can identify physically interesting events.
In this way, the HLT allows for storing only relevant events or parts thereof in order to save storage capacity.
On top of that the HLT performs a data compression, optionally assisted by the information obtained in the event reconstruction, to further reduce the amount of data.
Finally, the HLT can tag and classify the events during the online processing to support posterior processing steps.
In the second processing step, the ALICE \textbf{Offline} group performs a posterior, more comprehensive data analysis of all events that have been stored.

ALICE has subdetectors which require calibration in order to enable a proper event reconstruction.
Environment variables such as temperature and pressure affect the calibration.
These conditions change during a run which means the detectors must be continuously recalibrated.
There are several reasons that suggest performing online calibration in the HLT:
\begin{compactitem}
\item Without online calibration, HLT reconstruction has to use outdated calibration data from previous runs.
    Hence, the reconstruction will suffer from improper calibration.
\item \looseness=-1 The Offline reconstruction runs in several passes.
    The first passes are required for detector calibration.
    Performing at least the first calibration phase online in the HLT could make this Offline step obsolete and in this way save significant Offline compute resources.
\item The plans for LHC Run III include a significant increase in luminosity and ALICE plans a detector upgrade which will allow for the read out of considerably more data.
    Using the current data processing scheme, the increased amount of data will exceed the storage capacity by far and thus put higher demands on triggering, data compression, and online reconstruction capabilities.
    At this point in time, online calibration will become mandatory.
    In order to demonstrate the feasibility and gain experience, it is important to set up an online calibration scheme already now and take it to production mode, even if it is only a prototype for future detector upgrades.
\end{compactitem}

\looseness=-1
As explained above, a precise reconstruction needs up to date calibration data.
In the other way around, the calibration algorithm uses input from the reconstruction.
This imposes a cyclic dependency.
Therefore, a discussion of online calibration must also involve the online reconstruction.
The latter must provide sufficient data per time interval to enable online calibration.

This paper presents an analysis of the ALICE HLT data transport framework and the relevant reconstruction algorithms with respect to online calibration.
Based on experience gained during LHC Run I, we collect all the demands for online calibration and we identify shortcomings in the HLT during Run I which would prevent online calibration.
We present required components developed for the ALICE HLT in Run II, which enable a fast event reconstruction and necessary features for online calibration.
The requirements are formulated in an abstract way not related to online calibration in order to develop standalone solutions that can be used for other purposes as well.
The remainder of this paper is organized as follows.
Section~\ref{sec:terminology} introduces the current HLT scheme and all relevant terminology.
Thereafter, section~\ref{sec:requirements} collects a list of requirements for online calibration.
These requirements can be split in two parts: event reconstruction and framework requirements.
Sections~\ref{sec:tracking} and~\ref{sec:framework} cover our solutions for these two topics one after another.
Finally, section~\ref{sec:conclusions} summarizes the results and presents our conclusions and future plans.

\section{The ALICE HLT, its Data Transport Framework, and Track Reconstruction}
\label{sec:terminology}

\begin{figure}[h]
\includegraphics[width=0.68\textwidth]{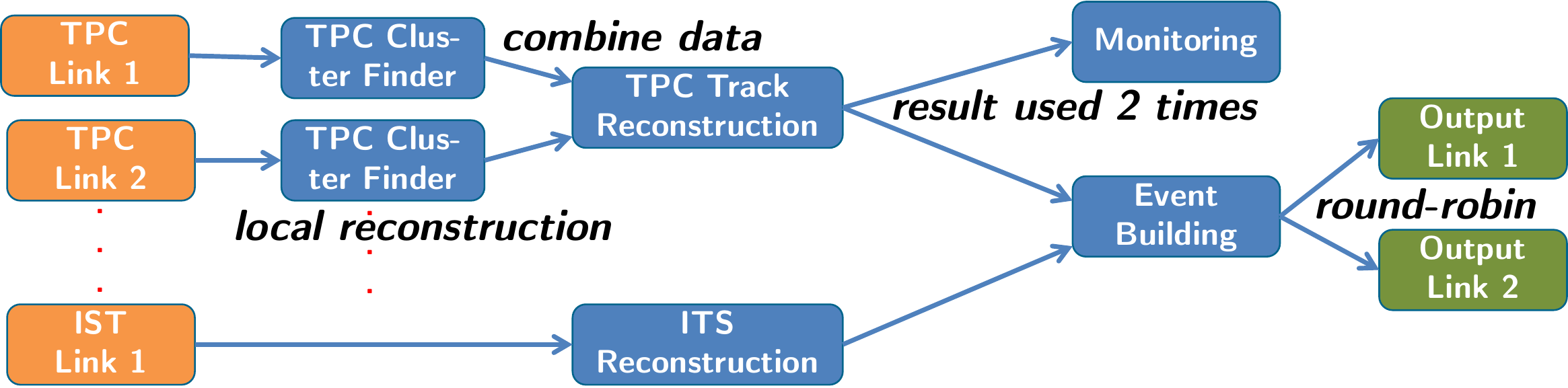}\hspace{2pc}%
\begin{minipage}[b]{0.26\textwidth}\caption{Simplified illustration of a reconstruction chain using the ALICE HLT framework.}
\label{fig:hltchain}
\end{minipage}
\end{figure}

The HLT employs a data transport framework based on the publisher-subscriber-principle.
Reconstruction is performed in a processing chain that consists of components.
Each component subscribes to other components in order to obtain the data published by them.
Components can reside on different compute nodes.
The data transport framework will automatically take care of all the necessary data transfer and scheduling.
Naturally, it makes sense to configure the processing chain in a way that minimizes such data transfer between nodes.
\fig{hltchain} presents a simple exemplary processing chain that demonstrates most of the relevant features.
There are three types of components:
Sources (in orange) do not subscribe to other components but only publish data received from detector links.
Sinks (in green) do not publish data for other components but send the data to \textbf{DAQ} (Data Acquisition) to store it.
Processing components (in blue) sit in between and perform the event reconstruction.
The scheme usually starts with local reconstruction (like the TPC Cluster Finding) which can operate on the data of only one detector link, \ie the component does not see the entire event but only a part thereof.
In later stages, data coming in from different detector links for the same event are merged together for global reconstruction, \ie the TPC Track Reconstruction subscribes to the data of all TPC links and the Event Building finally merges all data of one event.
Two components (like the Monitoring and Event Building in \fig{hltchain}) can subscribe to the same component if they need to operate on the same data.
In that case, both components receive the data for all events.
Alternatively, multiple instances of the same component can subscribe to one component such, that the events are send to them in a round-robin fashion.
This is important for load-balancing when bandwidth or processing capabilities of one component are insufficient.
\fig{hltchain} shows such an example with two Output Link components, one processing even-numbered and one processing odd-numbered events.
In reality, the chain is much more parallel and the round robin distribution is used at the very beginning (which is not shown in the figure for simplicity).
In this scheme, each detector link sees a part of every event, while the output links see all data of one event, but not every event.
Events flow through the chain from the left to the right in the figure and the framework operates like a huge pipeline:
Many events can be processed inside the chain simultaneously.
Hence, the HLT can be seen as a directed graph that must be without loops by construction.
(A loop would incur a cyclic data dependency that cannot be resolved.)

The most compute-intense step during event reconstruction, is the reconstruction of particle trajectories called \textbf{tracking}.
ALICE' main detector for this purpose is the TPC.
The HLT employs a GPU-accelerated implementation for TPC track reconstruction~\cite{bib:tns,bib:chep2012,bib:cnna}.

\section{Requirements for Online Calibration}
\label{sec:requirements}

\begin{figure}[h]
\includegraphics[width=0.57\textwidth]{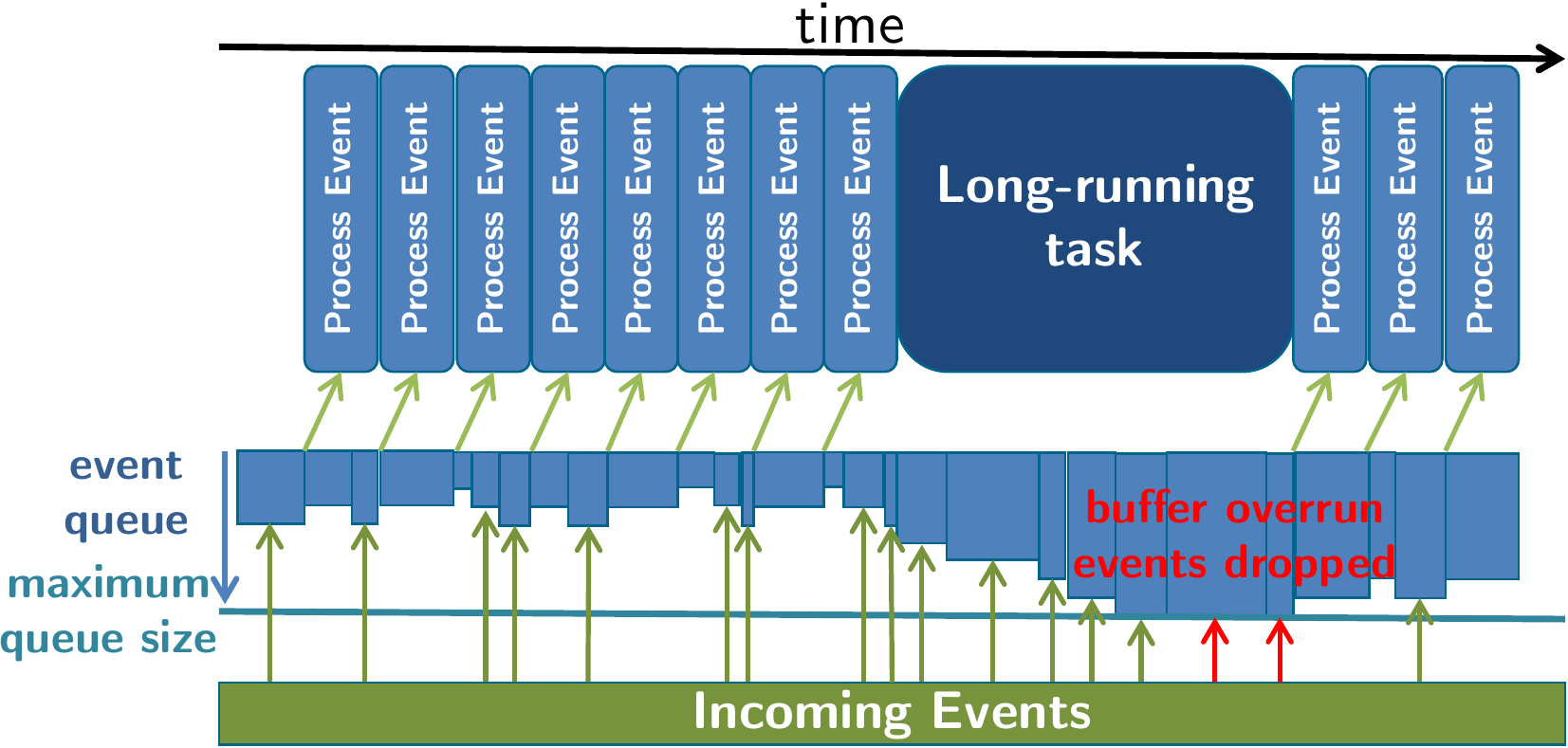}\hspace{2pc}%
\begin{minipage}[b]{0.36\textwidth}\caption{\label{fig:async1}Time development of the event queue buffer, which overflows during long-running tasks.}
\end{minipage}
\end{figure}

\begin{compactenum}
\item One step in the calibration is the matching of ITS and TPC tracks in order to obtain the TPC drift time.
  This step runs on top of the reconstruction, which makes \textbf{online track reconstruction} one requirement for online calibration.
\item The calibration uses several environment variables from external sensors like temperature and pressure.
  Data fed into the HLT chain must have an event ID and come from a detector link, which does not work \eg for sensor readings.
  In addition, for a reproducible calibration additional inputs such as sensor readings should reach the components synchronously with an event.
  This ensures that it is clear which sensor data was used during the processing of which event in the calibration.
  This imposes the requirement for \textbf{event-synchronous custom input sources} in an HLT chain.
\item In order for the HLT to exploit the calibration data it produces, these data must be fed back in the cluster transformation and the reconstruction respectively.
  This imposes a cyclic dependency: the calibration needs data from the reconstruction but it shall also send data to the reconstruction.
  This is prohibited by the HLT framework.
  One alternative method discussed in the initial planing of the HLT was the usage of the HCDB object storage on the file system, but tests during Run I demonstrated that this method will be too slow to work during data-taking.
  Therefore, the HLT needs a \textbf{fast method to feed back data in the chain}.
\item Components in the HLT reconstruct one event after another.
  Besides the size of the event, there is no difference in the treatment of them.
  This is different in the calibration, which builds calibration data based on a multitude of events.
  The calibration might require processing components, that perform long-running tasks from time to time (\eg a component that accumulates some data for every event but then runs a fit every n$^{th}$ event).
  During this long-running task, the event queue buffer might run full, as illustrated in~\fig{async1}.
  Hence, even though the average processing speed per event of that component might be fast enough, this may still lead to loosing events.
  Therefore, online calibration requires a feature to \textbf{run long-running tasks asynchronously}.
\end{compactenum}

This means that apart from the online reconstruction, the HLT must provide three features to enable online calibration during Run II.
The HLT data transport framework has proven its stability during Run I and we want to keep the framework as is where possible.
The new features can and are formulated in an abstract way and the implementations presented below show how they can be provided in standalone processing components.
This has the advantage that they do not require modifications to the framework and they can be used in other places as well besides online calibration.

\section{GPU based Track Reconstruction with OpenCL}
\label{sec:tracking}

\subsection{Common Source Code}
A problem with many GPU implementations is that one wants to keep a CPU code for reference.
Maintaining two different code bases and in particular keeping them in sync requires significant effort.
Therefore, the HLT tracker bases on a common source code approach.
All code belonging to the actual algorithm is contained in a generic source code file.
Wrappers for the different \textbf{APIs} (\fBold{Application} \fBold{Programming} \fBold{Interface}) include this common file and contain all API, device, and runtime specific code.
Listings~\ref{lst:common} --~\ref{lst:cpu} show a simplified example how this works for the fit of a larger number of tracks.
Listing~\ref{lst:common} contains the actual fit algorithm, Listing~\ref{lst:cuda} is the GPU wrapper that uses the CUDA API to define and launch a kernel, and Listing~\ref{lst:cpu} shows the CPU wrapper, that can optionally use OpenMP for multithreading on the processor.
In case of the tracker, 95\% of the code is in the common generic files (common.cpp in the example).
Changes to the algorithm must be done only once which ensures good maintainability.

\noindent
\begin{minipage}[][][t]{.29\textwidth}\vspace{0pt}
\begin{lstlisting}[label={lst:common},frame=single,basicstyle=\ttfamily \tiny,caption={common.cpp}]{Name}
__DECL FitTrack(int n) {
...
}


//Generic file
//Included by wrapper_cuda.cu
//and wrapper_cpu.cpp
\end{lstlisting}
\vfill
\end{minipage}
\hfill
\begin{minipage}[!t]{.33\textwidth}
\begin{lstlisting}[label={lst:cuda},frame=single,basicstyle=\ttfamily \tiny,caption={wrapper\_cuda.cu}]{Name}
#define __DECL __device void
#include "common.cpp"
__global void FitTracksGPU() {
  FitTrack(threadIdx.x);
}
void FitTracks() {
  FitTracksGPU<<<nTr>>>();
}
\end{lstlisting}
\end{minipage}
\hfill
\begin{minipage}[!t]{.31\textwidth}
\begin{lstlisting}[label={lst:cpu},frame=single,basicstyle=\ttfamily \tiny,caption={wrapper\_cpu.cpp}]{Name}
#define __DECL void
#include "common.cpp"
void FitTracks() {
#pragma omp parallel for
  for (int i = 0;i < nTr;i++) {
    FitTrack(n);
  }
}
\end{lstlisting}
\end{minipage}
\subsection{GPU Tracking with OpenCL}

\looseness=-1
The original HLT TPC GPU tracker based on the NVIDIA CUDA API, which was at that point the only suitable API due to its \CC{} support.
However, relying exclusively on CUDA imposes a very tight vendor lock for the HLT, which in this case must use NVIDIA GPUs.
In order to become vendor independent, we have created a wrapper for OpenCL, which is a free, open, modern API for GPU programming similar to CUDA.
In the same way as for the CPU and for CUDA, the OpenCL tracker is implemented via a wrapper that includes the shared algorithm code.
The shared code was adapted where needed to work within the OpenCL framework.
Basing on the standard ALICE analysis framework AliRoot that is written in \CC{}, the GPU tracker requires a \CC{} capable framework.
OpenCL itself does not support the \CC{} language for compute kernels.
However, OpenCL enables the specification of vendor extensions and AMD provides an adequate OpenCL \CC{} kernel language extension.
Currently, this imposes the restriction that the OpenCL GPU tracker can only run on AMD GPUs.
However, as CUDA is still supported, the HLT has now the option to employ either CPUs or NVIDIA GPUs or AMD GPUs for tracking.
This flexibility is likely to grow as soon as other vendors provide OpenCL \CC{} extensions as well or when \CC{} becomes supported by a newer revision of the OpenCL specifications.

The main challenge for the adoption of OpenCL turned out to be pointers to different memory spaces on the GPU.
OpenCL distinguishes pointers to global GPU memory (\textit{\_\_global}), local GPU memory (\textit{\_\_local}), constant GPU memory (\textit{\_\_constant}) and private GPU memory.
A conversion of pointers to a different address space is illegal.
Listing~\ref{lst:openclpointers} shows some examples.

\noindent
\begin{minipage}[!t]{.61\textwidth}
\begin{lstlisting}[label={lst:openclpointers},frame=single,basicstyle=\ttfamily \tiny,caption={Pointers to different address spaces in OpenCL}]{Name}
__kernel void test (__global int* a /*Pointer to global space*/) {
  int b;                   //Variable in private space
  int* c = &b;             //Pointer to variable in private space
  c = a;                   //ILLEGAL (global to private cast)
  __local int d;           //Variable in local space
  __local int e = &d;      //Pointer to variable in local space
  e = a;                   //ILLEGAL
  e = &b;                  //ILLEGAL
}
\end{lstlisting}
\end{minipage}
\hfill
\begin{minipage}[!t]{.33\textwidth}
\begin{lstlisting}[label={lst:openclfail},frame=single,basicstyle=\ttfamily \tiny,caption={OpenCL Problem.}]{Name}
class a {public:
  const __global int* GetB() {
    return(&b);
  }
  private: int b;
};
__kernel void foo(__global a* tmp) {
  const __global int* b = tmp.GetB();
}
\end{lstlisting}
\end{minipage}

\looseness=-1
The tracker code employs plenty of objects with member functions that return pointers to the object's memory.
If the object resides in global address space, the return type of such a member function must be declared as a global address space pointer.
However, in principle, the object can reside in different address spaces.
And indeed, during track reconstruction such objects are moved \eg from global to local memory for performance reasons.
However, the member function's return type must be declared with one particular address space qualifier.
The compiler compiles all member functions multiple times, once for the object residing in each address space.
Unfortunately, the compiler only varies the object's address space itself, \ie the space where its members are stored, but not the return type of member functions.
Therefore, the code in Listing~\ref{lst:openclfail} fails to compile, because when it is compiled for an object in local address space the function {\ttfamily \small GetB} would cast a local pointer to a global pointer.
Many problems like this exist in the tracker code, not restricted to return types but also to function parameters and pointers in general.
One possible solution for pointers in function parameters are templates, which avoid the above compilation error because in \CC{} substitution failure is not an error.
Unfortunately, this does not work for the return type of functions because functions differing only by the return type cannot overload each other.
A template type which is used exclusively in the return type must be explicitly defined at every function call, which would make the code much more complex and difficult to maintain.
(The call in Listing~\ref{lst:openclfail} would become {\ttfamily \small tmp.GetB<\_\_global int*>();}.)

Instead, we use a different approach where the object knows its own memory address space via a template argument such that return types can be declared properly.
Listing~\ref{lst:openclsolution} presents our solution to this.
The declaration {\ttfamily \small \_\_local A<Local>} declares an object of class A that knows it resides in local address space.
Finally, the only required code changes are in the declarations of member functions.
The definition of objects already had to use a macro before to pass the correct keyword (like {\ttfamily \small \_\_local}) to the compiler for the different APIs like CUDA and OpenCL.
This macro was simply adapted.

\begin{lstlisting}[label={lst:openclsolution},frame=single,basicstyle=\ttfamily \tiny,caption={Template solution for OpenCL address type qualifiers}]{Name}
enum LocalOrGlobal { Local, Global };
template<LocalOrGlobal, typename L, typename G> struct MakeType;
template<typename L, typename G> struct MakeType<Local, L, G> { typedef L type; };
template<typename L, typename G> struct MakeType<Global, L, G> { typedef G type; };
template<LocalOrGlobal LG> struct A {
  float x[10];
  typename MakeType<LG, __local float *, __global float *>::type foo() { return x; }
};
__kernel void foo(__global A<Global> * ptr) {
  __global float* aa = ptr->foo();
  __local A<Local> obj;
  __local float* aaa = obj.foo();
}
\end{lstlisting}

The tracking time on current NVIDIA and AMD GPUs using the two different APIs is similar.
Table~\ref{tab:perf} gives an overview.
Note that the NVIDIA Maxwell GPU family was released only after the new HLT cluster was built.
It was thus no option for the cluster and is listed only for reference.
We have chosen the AMD FirePro S9000 GPU for the new HLT cluster and we have installed 160 such cards in the HLT for Run II.
Luckily, we had access to several hundred AMD FirePro GPUs during cluster commissioning of the Sanam cluster at GSI, where we could run the OpenCL tracker as a stress test.
This test confirmed the stability of our solution.
A comparable traditional system based on CPUs that delivers the same tracking throughput as our solution would need more than 100 additional servers.
Hence, the GPU tracker results in a significant reduction of acquisition costs and power consumption of the HLT.
On top of the speedup shown in Table~\ref{tab:perf}, we have started to adopt new GPU features, in particular the possibility to run multiple different kernels in parallel.
We see already a 15\% performance improvement for the next tracker version and expect to achieve more than 30\% in the end.

\begin{table}[h]
\caption{\label{tab:perf}Performance of the GPU tracker on different GPUs}
\begin{center}
\begin{tabular}{lrr}
\br
GPU Model & Specifications & Tracking time\\
\mr
NVIDIA GTX480 (Fermi) (was used in Run I) & 448 shaders, 1215 MHz & 174 ms\\
NVIDIA GTX780 (Kepler) & 2304 shaders, 863 MHz & 155 ms\\
NVIDIA Titan (Kepler) & 2688 shaders, 837 MHz & 146 ms\\
\textbf{AMD S9000 (Tahiti)} (will be used in Run II) & 1792 shaders, 900 MHz & \textbf{145 ms}\\
NVIDIA GTX980 (Maxwell) & 2048 shaders, 1126 MHz & 120 ms \\
\br
\end{tabular}
\end{center}
\end{table}

\section{New Features of the ALICE HLT Framework}
\label{sec:framework}

\subsection{Asynchronous Processing}
\label{sec:async}

\begin{figure}[h]
\includegraphics[width=0.56\textwidth]{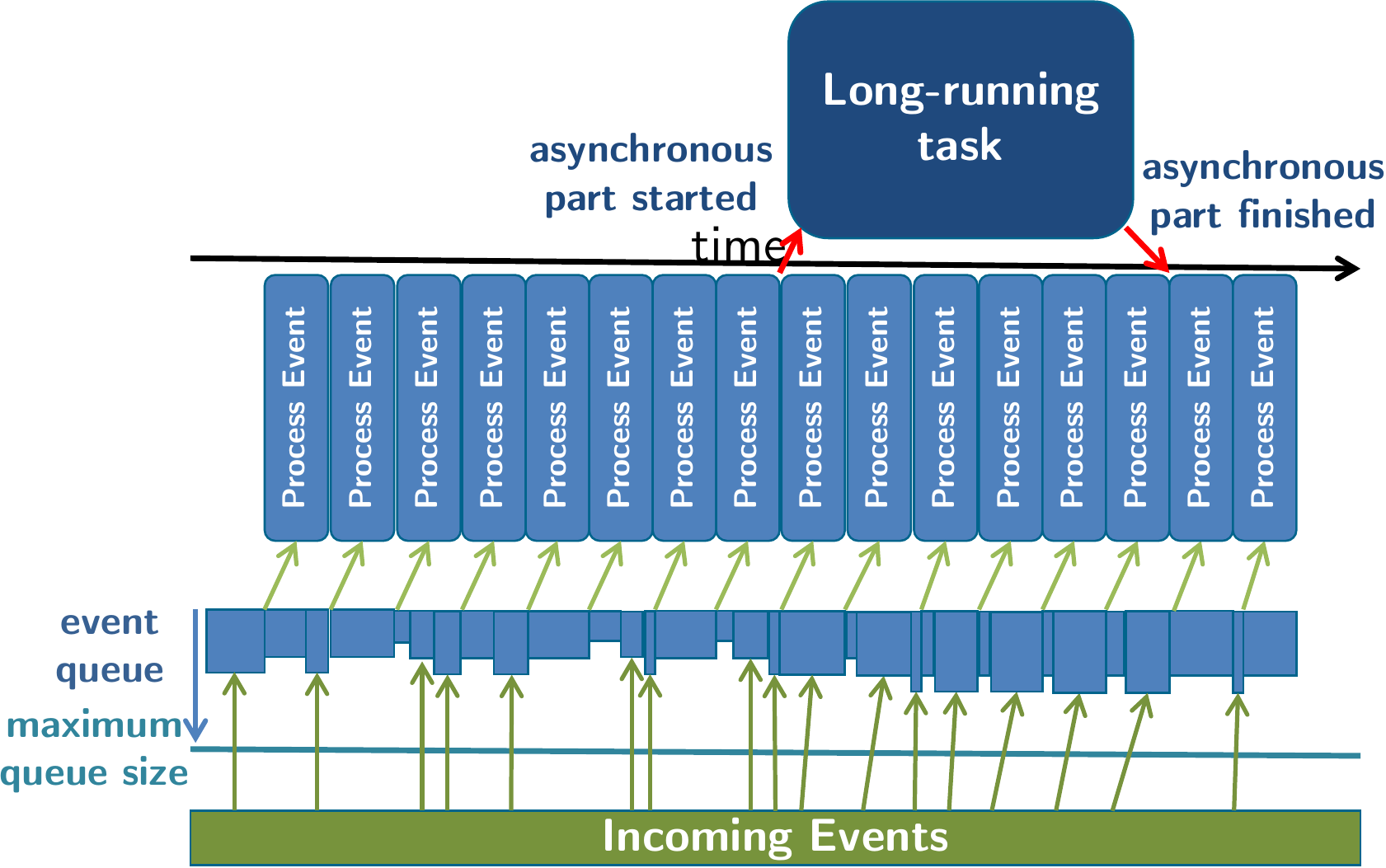}\hspace{2pc}%
\begin{minipage}[b]{0.38\textwidth}\caption{\label{fig:async2}Long-running task processed asynchronously. Event queue does not run full.}
\end{minipage}
\end{figure}

\looseness=-1
We have created a new special framework processing component base-class that supports an interface for offloading asynchronous subtasks.
It spawns a thread and provides an interface for initialization, queuing and synchronization of long-running tasks, as well as passing ROOT objects and pointers to memory forth an back.
The component starts the asynchronous task when processing one event, it checks every event whether the asynchronous job has finished, and as soon as it has, it can use the result in the standard event processing loop.
\fig{async2} illustrates the process.
Event processing continues during the long-running task such that the buffer does not run full.

\subsection{Custom Data Sources}
\label{sec:source}

We realize the custom data sources via an intermediate component called event trigger.
Components receiving data from the detector links send empty packages without payload to this event trigger.
The event trigger receives only the information that an event was received, and basic information about this event such as the event number.
Then, it passes this information to custom data source components and triggers the input of custom data such as sensor values or calibration results.
This ensures that the custom data is fed into the chain synchronously with an event, and it allows the HLT to keep track globally at what event number the custom source provided an updated value.
\fig{framework2} at the end of this section illustrates an HLT chain that contains the new features including the event trigger and custom data sources.

\subsection{Feedback Look}

The HLT will provide the Feedback Loop capability via two components that send and receive data via the ZeroMQ message queue framework outside of the standard data transport framework.
These components rely on the asynchronous processing features presented in section~\ref{sec:async} such that they do not block the dataflow.
And the receiving component uses the custom data source feature of section~\ref{sec:source}, to feed the data into the chain synchronously.
This approach shows the usefulness of the standalone component approach, which are not limited to online calibration but are reused here.
Because updated calibration data will be fed into the chain via the custom data source component described above, the HLT can keep track at what event number the calibration was updated.
We will the calibration objects themselves and a reference of the calibration object used together with an event to ensure reproducibility.

\subsection{Full new HLT scheme}

\begin{figure}[h]
\includegraphics[width=0.65\textwidth]{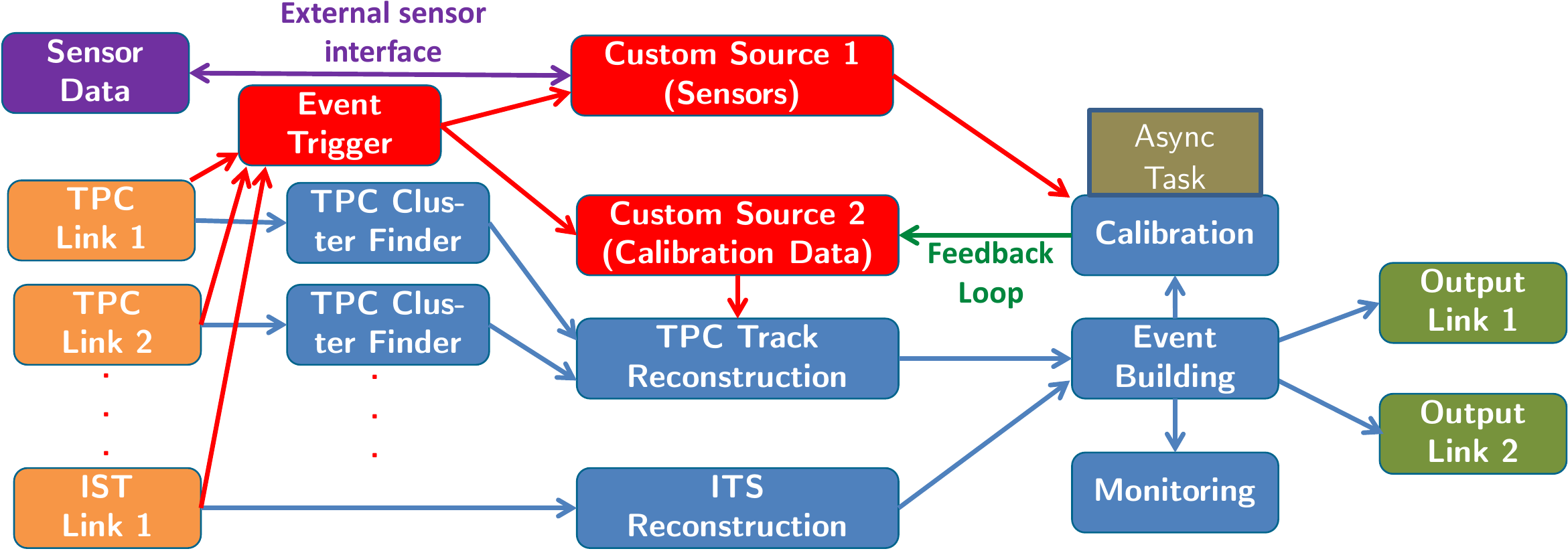}\hspace{2pc}%
\begin{minipage}[b]{0.3\textwidth}\caption{\label{fig:framework2}HLT reconstruction chain with online calibration.}
\end{minipage}
\end{figure}

\fig{framework2} illustrates how everything is supposed to work together in the HLT in Run II as soon as all components, in particular the calibration itself, have been finished.
The GPU based tracking provides real-time event reconstruction, the calibration component utilizes the asynchronous processing feature to create calibration objects, these objects are fed back into the chain via the feedback loop and the custom data source component, and finally the reconstruction can synchronously switch to the new calibration object as soon as it is available.

\section{Conclusions and Next Steps}
\label{sec:conclusions}

We have adapted the GPU based HLT TPC track reconstruction to optionally use OpenCL, which makes the HLT vendor independent.
160 AMD FirePro GPUs have been deployed at the new HLT for Run II and the stability of the system was shown.
With these GPUs, the HLT can provide online TPC tracking at the maximum expected rates during Run II.

\looseness=-1
We have identified three missing components in the HLT framework that are required for online calibration and we have presented solutions for them.
The integration is nearly finished.
In parallel, we have started the development of a calibration component, which will base on all the work presented in this paper.
After combining all these results, we hope to commission the online calibration during Run II, based on the already deployed event reconstruction with tracking.

\ack{This work has been partially funded by HIC for FAIR.}

\end{document}